\documentclass[twocolumn,preprintnumbers,floatfix,prb,groupaddress]{revtex4}
\usepackage{amsmath}
\usepackage{amssymb}
\usepackage{graphicx}% Include figure files
\usepackage{dcolumn} % Align table columns on decimal point
\usepackage{ragged2e} % to justify the captions
\usepackage{bm}
\usepackage{epsfig}
\usepackage{hyperref}
\usepackage{xcolor}

\newcommand{\bd}{\mathbf{d}}
\begin{document} 
\title{Excitonic magnetism in $d^6$ perovskites}

\author{J. Fern\'andez Afonso}
\affiliation{Institute of  Solid State Physics, TU Wien, Wiedner Hauptstr. 8, 1040 Wien, Austria}
\affiliation{Institute of Physics, Academy of Sciences of the Czech Republic, Cukrovarnick\'a 10,
Praha 6, 162 53, Czech Republic}

\author{J. Kune\v{s}}
\email{kunes@ifp.tuwien.ac.at}
\affiliation{Institute of  Solid State Physics, TU Wien, Wiedner Hauptstr. 8, 1040 Wien, Austria}
\affiliation{Institute of Physics, Academy of Sciences of the Czech Republic, Cukrovarnick\'a 10,
Praha 6, 162 53, Czech Republic}

\date{\today}

\begin{abstract}

We use the LDA+U method to study the possibility of exciton condensation in perovskites of transition metals with $d^6$ 
electronic configuration such as LaCoO$_3$. For realistic interaction parameters 
we find several distinct solutions exhibiting spin-triplet exciton condensate, which 
gives rise to a local spin density distribution while the ordered moments are vanishingly small.
Rhombohedral distortion from the ideal cubic structure suppresses the ordered state, 
contrary to the spin-orbit coupling which enhances the excitonic condensation energy. 
We explain the trends observed in the numerical simulations
with the help of  a simplified strong-coupling model. Our results indicate that LaCoO$_3$ is close to the 
excitonic instability and suggest ways to make it order.

\end{abstract}

\maketitle

%%%%%%%%%%%%%%%%%%%%%%%%%%%%%%%%%%%%%%%%%%%%%%%%%%%%%%%%%

\section{Introduction}\label{intro}
Long-range order (LRO) is a ubiquitous phenomenon in  
correlated materials. Mott insulators built from atoms with partially filled $d$- or $f$-shells
and degenerate ground states are a typical example. The tendency to LRO is not
restricted to strongly correlated Mott insulators, but often extends to moderately correlated metals
in BEC-BCS-like fashion.

%For lighter elements this typically amounts to spin ordering driven by isotropic exchange or orbital ordering with anisotropic exchange. 
%For heavier elements
%the spin-orbit coupling forms tightly bound spin-orbitals and thus gives rise to generally anisotropic exchange.

Having no local degrees of freedom, materials with singlet atomic ground states are unlikely to exhibit LRO. 
Existence of low-energy atomic excitations, e.g., in materials close to the spin-state (high-spin--low-spin) crossover, 
may change this conclusion. Taking the inter-atomic interactions into account in such a case, 
a lattice in an all-singlet state is not necessarily the global ground state. Magnetic order
on top of the excited multiplets or spin-state order, i.e., a lattice decorated 
with several atomic multiplets, may have a lower energy. 

\begin{figure}
\includegraphics[width=\columnwidth]{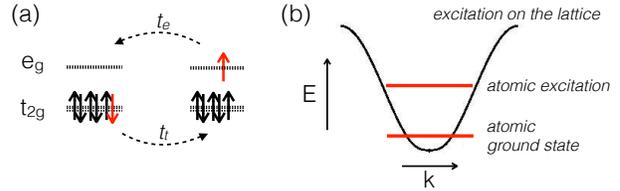}
\caption{(a) A cartoon view of the exciton propagation in $d^6$ cobaltites. (b) The energy level scheme of a system
unstable to exciton condensation. The atomic levels (red) and the dispersion of the atomic excitation on lattice due to
the hopping process (a).}
\label{fig:hop}
\end{figure}

Yet another interesting possibility arises from a non-negligible amplitude to exchange
different multiplets on neighboring atoms. This is the case when the atomic multiplets 'differ' by an
electron-hole pair and thus excited state can be viewed as an exciton added to the atomic ground state.
The multiplet inter-atomic exchange is then equal to hopping, which endows such an exciton with 
dispersion. If the bottom of the dispersion falls below the all-singlet energy the system is unstable
towards exciton condensation~\cite{mott61}, i.e., Bose-Einstein condensation of excitons
into the lowest energy state. (See Fig.~\ref{fig:hop}.) 
The ferromagnetic Hund's coupling favors spinful ($S=1$ for weak spin-orbit coupling) low-energy excitons,
condensation of which involves breaking of the spin-rotation symmetry,
over the spinless ($S$=0) excitons.

LRO of this type, also called excitonic magnet, has been studied in simple lattice 
models~\cite{kha13,kunes14a,kunes14b,kaneko14,kaneko15,chaloupka16} and proposed
to take place in $4d^4$ materials such as Ca$_2$RuO$_4$~\cite{kha13,jain15} or $3d^6$ cubic materials such as the
LaCoO$_3$ family~\cite{kunes14b}. Here we focus on the latter.

Although no thermally driven phase transition has been observed in LaCoO$_3$, a proximity to an ordering instability 
or short range correlations have been discussed for a long time. 
The most studied proposals include the spin state ordering,~\cite{bari72,knizek09,kunes11} cooperative
Jahn-Teller effect~\cite{korotin96,yamaguchi97} and magnetic correlations.~\cite{tokura98,zhang12}  These states can be viewed as a classical LRO of
atomic states/multiplets and are generally expected to be accompanied by Co-O bond length disproportionation.
Despite experimental effort the Co-O bond length disproportionation remains controversial.~\cite{radaelli02,maris03,pandey06,sikolenko16}
Recently, high magnetic field experiments revealed a meta-magnetic transition with a peculiar temperature dependence.~\cite{altarawneh12,rotter14,ikeda16a}
Field-induced exciton condensation was suggested as an explanation~\cite{ikeda16a,sotnikov16}
and studied in a simplified two-orbital Hubbard model by means of an effective strong coupling model~\cite{ishihara16a,tatsuno16} and the 
dynamical mean-field theory.~\cite{sotnikov16}
Exciton condensation was also proposed~\cite{kunes14b,kazuma17} to explain the phase transition observed in the Pr$_{0.5}$Ca$_{0.5}$CoO$_3$ 
family.~\cite{tsubouchi02,hejtmanek10,knizek13,ikeda16b}

Unlike the other LROs discussed above, exciton condensation, which involves a spontaneous coherence between different atomic 
multiplets on the same ion, does not have a classical analog. 
The spin and orbital degeneracy of the excited multiplet allows numerous distinct exciton condensates, i.e. states that
cannot be transformed to one another by an operation from the Hamiltonian symmetry group. These distinct condensates
may exhibit quite different properties. States that possess or lack ordered atomic moments~\cite{kunes14c} or time-reversal
symmetry~\cite{halperin68,kunes16} can be built from spinful ($S=1$) excitons. The orbital degeneracy present in $d^6$ perovskite
makes the condensate 'zoo' even richer. The driving force behind the exciton condensation, the exciton hopping in Fig.~\ref{fig:hop}a,
does not distinguish between the condensate species. There are the remaining inter-atomic interactions which
select the most stable condensate. The shear number of channels for these interactions makes their realistic modeling questionable.

Therefore we resort to an approximate static mean-field treatment with minimum adjustable parameters embodied in
the LDA+U approach~\cite{ldau}. The method  is too simple to capture quantum or thermal fluctuations and
 thus cannot be used to study the transition to the condensed phase.
On the other hand, it can capture the exciton condensate and thus allows us to answer the following basic questions.
Are there ordered solutions at all and how does their stability depend on the strength of the on-site Coulomb interaction?
Which of the many possible excitonic states has the lowest energy and why? How do the ordered solutions respond
to the lattice distortions and inclusion of the spin-orbit coupling (SOC)?

The paper is organized follows. In section~\ref{sec:comp} we introduce the multi-component order parameter describing the exciton
condensate and summarize the computational parameters used in this study. In section~\ref{sec:results} we present the obtained
ordered solutions. We have performed extensive calculations for the hypothetical cubic systems without SOC varying the 
interaction parameters and initial conditions. Subsequently, we investigate the effect of the rhombohedral distortion, present in the real LaCoO$_3$ structure, and SOC for a selected set interaction parameters. The main features of our numerical results are summarised in section
\ref{sec:discussion}. We introduce two strong-coupling models with different degree of simplification, which allow
us to present semi-quantitative explanations of the observed trends. The key observations are them summarised in the compact
form in section~\ref{sec:conclusions}.

\begin{figure}
\includegraphics[width=\columnwidth]{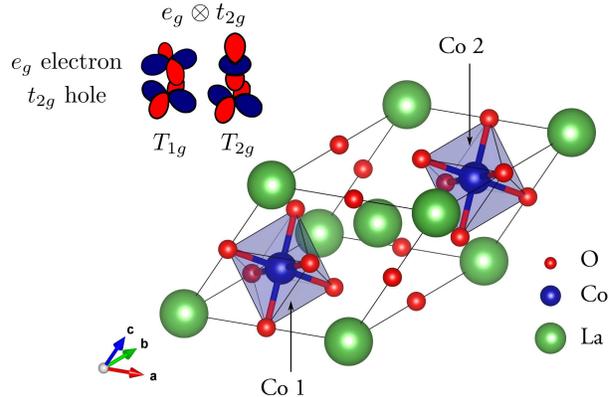}
\caption{The unit cell used in the calculations. In the upper left corner we show the orbital structure of the $T_{1g}$ and $T_{2g}$ excitons.}
\label{fig:struct}
\end{figure}

\section{Computational method}\label{sec:comp}
The calculations reported here were performed in the framework of the density functional theory using the LDA+U approach,~\cite{ldau}
with the 'U' interaction in the $3d$ shell of Co and the double-counting
correction corresponding to so called fully localized limit.~\cite{fll} The spin polarization entered only the orbitally dependent potential
acting on the $3d$ states of Co while the LDA exchange-correlation potential was not spin polarized. The calculations were done with the Wien2k~\cite{wien2k}
software. In all cases we used the unit cell of Fig.~\ref{fig:struct}, which can accommodate the real LaCoO$_3$ as well as the staggered excitonic order, 
which was found to have lower total energy than a uniform one. No orbital or spin point symmetry was assumed. 

Several crystal structures were used in this study: hypothetical cubic perovskite structure with $a$=3.8498~\AA, the experimental 
LaCoO$_3$~\endnote{from Ref.~\onlinecite{radaelli02} at 5~K} and a series of structures interpolating between the two.
The calculations were performed with the $10 \times 10 \times 10$ k-mesh in the full Brillouin zone, muffin-tin radii (in bohr): 2.5 for La, 1.99 for Co and 1.92 for O,
and  the plane wave cut-off $R_{mt}K_{max} = 6$. The Co $3d$ occupation matrices discussed in the following are
expressed in the local orbital and spin coordinates tied to the CoO$_6$ octahedra--axes pointing (approximately) along the Co-O bonds.~\endnote{Rotation
of the spin coordinates required small modification of the LAPWDM code from Wien2k package.}
%In order to study the existence the excitons later described we have define the local coordinates of the cobalt according with the octahedral environment CoO$_6$. 
%The unit cell used for the cubic structure is shown in Fig.\~ref{fig:struct} left. It is formed by two formula units which allows the antiferro-ordering between nn's Co. The %corresponding lattice parameters were set to $a$=5.43~\AA  and $\alpha = \pi/3$ rad.

\subsection{Excitonic order parameter}\label{EC_lacoo3}
Although the present method is a fermionic mean-field theory, we find it instructive to refer to the strong-coupling 
limit of the present problem, which can be formulated in terms of hard-core bosons.~\cite{balents00,kunes15}
The lattice occupied by atoms in the low-spin state can then be viewed as the bosonic vacuum.
A bosonic condensate is, in general, a superposition of the vacuum with some other bosonic states. 
We consider a wave function of the condensed system in the form of a product over Co sites 
\begin{equation}
\label{eq:wf}
\left| \Psi  \right\rangle = \prod\limits_k \left(s_k \left| LS \right\rangle_k  + \xi^\alpha_k\left| IS_{\alpha} \right\rangle_k  + \zeta^{\beta}_k \left| HS_{\beta} \right\rangle_k \right).
\end{equation}
Here, we have a superposition of the low-spin (LS, S=0), intermediate-spin (IS, S=1) and high-spin (HS, S=2) states on each Co site.
Evaluating the local occupation matrix in the $d$-shell of Co in the above state, anomalous matrix elements, violating the cubic and spin-rotational symmetry,
appear. These terms will serve as our order parameter.

We consider only IS states of the $T_{1g}$ symmetry which correspond to the  $d_{x^2 - y^2}\otimes d_{xy}$, $d_{z^2 - x^2}\otimes d_{xz}$ and 
$d_{y^2 - z^2}\otimes d_{yz}$ electron-hole excitations
(excitons) of the LS state. Only such excitons have sizeable amplitudes to propagate as composite objects and to condense.~\cite{sotnikov16}. 
This assumption will be justified {\it a posteriori} by unrestricted calculations, which show that the order parameter has indeed the $T_{1g}$ orbital symmetry.
The anomalous  elements of the Co-$3d$ occupation matrix can be described by 18 Hermitean operators
\begin{equation}\label{O_operators}
\begin{split}
\begin{array}{*{20}{c}}
\hat{O}'_{\alpha\beta}=\frac{1}{4}\sum\limits_{i,i'=1}^5\sum\limits_{\sigma\sigma'=\uparrow,\downarrow}(\Gamma'_{\alpha})_{ii'}(\tau_{\beta})_{\sigma\sigma'}\hat{c}^{\dagger}_{i\sigma}\hat{c}^{\phantom\dagger}_{i'\sigma'} \\[2ex]
\hat{O}''_{\alpha\beta}=\frac{1}{4}\sum\limits_{i,i'=1}^5\sum\limits_{\sigma\sigma'=\uparrow,\downarrow}(\Gamma''_{\alpha})_{ii'}(\tau_{\beta})_{\sigma\sigma'}\hat{c}^{\dagger}_{i\sigma}\hat{c}^{\phantom\dagger}_{i'\sigma'}.
\end{array}
\end{split}
\end{equation}
Here $\hat{c}^{\dagger}_{i\sigma}$ ($\hat{c}^{\phantom\dagger}_{i\sigma}$) are the creation (annihilation) operators for an electron in the Co $3d$ state with orbital and spin indices $i$ and $\sigma$ acting on the same atom. (The site indices are not shown for sake of simplicity).
The Pauli matrices  $\tau_{\beta}$ ($\beta=x,y,z$) capture the spin-triplet character of the excitonic order 
while the $5\times5$ $\Gamma_{\alpha}$ ($\alpha=\hat{x},\hat{y},\hat{z}$) matrices describe the orbital structure.
The primed  and double-primed $\Gamma_{\alpha}$ matrice correspond
to the symmetric (density-like) and anti-symmetric (current-like) combinations respectively. The explicit form of $\Gamma_{\alpha}$ in cubic harmonics
as well as spherical harmonics bases can be found in the Appendix.

We define the order parameter $\phi_{\alpha\beta}$ as a complex $3\times3$ matrix formed by the expectation values of the operators $\hat{O}'_{\alpha\beta}$ and $\hat{O}''_{\alpha\beta}$
\begin{equation}\label{phi_expect_O}
\phi_{\alpha\beta}=\langle \hat{O}'_{\alpha\beta} \rangle +i\langle \hat{O}''_{\alpha\beta} \rangle = \phi'_{\alpha\beta} + i\phi''_{\alpha\beta}.
\end{equation}
The columns $\boldsymbol{\phi}$ transform as a pseudovectors ($T_{1g}$ representation) under the $O_h$ point group. 
The rows of $\boldsymbol{\phi}$ transform as a vectors ($S=1$ representation) under the $SU(2)$ spin rotations. 
We further define the norm $|\boldsymbol{\phi}|=\sqrt{\sum_{\alpha\beta}|\phi_{\alpha\beta}|^2}$ and
introduce a special variable for the spin(row)-vectors $\boldsymbol{\eta}_{\alpha}$ ($\alpha=\hat{x},\hat{y},\hat{z}$),
$(\boldsymbol{\eta}_{\alpha})_{\beta}=\phi_{\alpha\beta}$.

Our choice of the order parameter can be tested {\it a posteriori} by comparing the calculated occupation matrices 
with the expected form of the anomalous part:
\begin{equation}\label{D_mm_ss}
%\begin{split}
\langle \hat{c}^{\dag}_{i \sigma} \hat{c}_{i'\sigma '} \rangle_A =\sum_{\alpha,\beta}
 \phi'_{\alpha\beta} (\Gamma'_\alpha)_{ii'}^* (\tau_\beta)_{\sigma \sigma '}^* + \phi''_{\alpha\beta} (\Gamma''_\alpha)^*_{ii'} (\tau_\beta)_{\sigma \sigma '}^*.
%\end{split}
\end{equation}
%where the $\langle\rangle_A$ subindex stands for the anomalous part that which violates the cubic and spin-rotational symmetry.

\section{Results}\label{sec:results}
We start our presentation with idealized cubic structure without SOC, which possesses the exact $O_h$ site symmetry and $SU(2)$ spin rotational symmetry,
and proceed by continuously switching the rhombohedral distortion and adding the spin-orbit coupling. 
\begin{figure}
\includegraphics[width=\columnwidth]{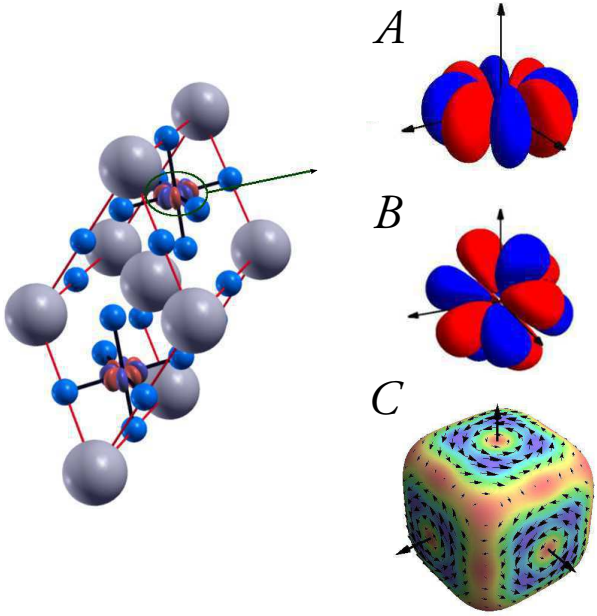}
\caption{\label{fig:laco1}Spin density distribution in the $A$, $B$ and $C$ condensate phases. Left, the isosurface of the spin density (red-positive, blue-negative) in the $A$-phase of ideal cubic structure obtained with Wien2k. Right, detail of the spin density distributions around the Co atom in the $A$, $B$ and $C$ states.
For the collinear ($A$, $B$) distribution we show the positive and negative isosurfaces of the spin projection. For the non-collinear ($C$) distribution we show the spin distribution on the
$x^4+y^4+z^4=\text{const}$ surface where the normal component vanishes. (Color codes the amplitude of the tangent component with direction marked by the arrows.) 
}
\label{phase_diag_UJ}
\end{figure} 

\begin{figure}
\includegraphics[width=\columnwidth]{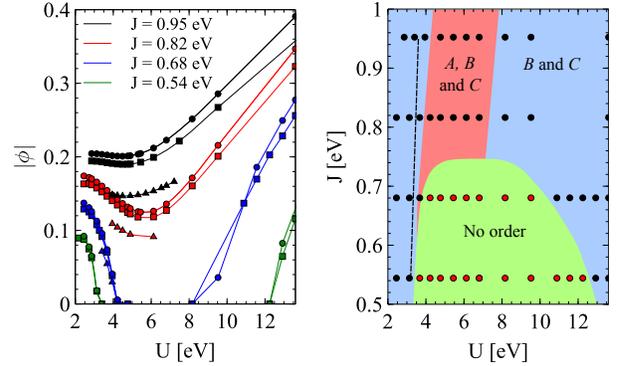}
\caption{(Left) The order parameter $|\phi|$ vs U for various J values. The $A$, $B$ and $C$ phases are represented with triangles, squares and circles. (Right) The $U$-$J$ phase diagram
indicating the local stability of the three excitonic phases. The circles mark the points where actual calculations were performed. The dashed line shows where the gap
opens in the normal phase.}
\label{phase_diag_UJ}
\end{figure} 
\begin{figure}
\includegraphics[width=\columnwidth]{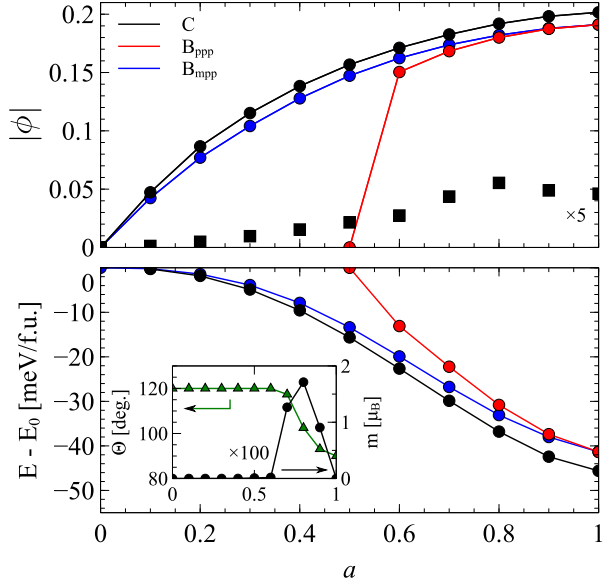}
\caption{(Upper panel) The evolution of the staggered real $|\phi'|$ (circles) and uniform imaginary $|\phi''|$ (squares) components of the order parameter along
the path between the cubic ($a=1$) and experimental ($a=0$) LaCoO$_3$ structures in the $C$, $B_{mpp}$ and $B_{ppp}$ phases. (Bottom panel)
The condensation energy - the difference between the total energy of ordered and normal solution for the three stable solutions. In the inset we show
the evolution of the ordered (ferromagnetic) moments on Co sites and the angle $\Theta$
between the spin vectors $\boldsymbol{\eta}_{\beta}$ and $\boldsymbol{\eta}_{\beta'}$ in different orbital channels in the $C$ phase.
The results were obtained with $U$=3.90~eV and $J$=0.95~eV.}
\label{path_ideal_real}
\end{figure}     
                                             
\subsection{Cubic structure without spin-orbit coupling}
Three distinct ordered states referred to as $A$, $B$ and $C$ were found to be stable for the cubic system without SOC. 
The $A$-type states are described by the order parameter~\endnote{This describes one of the set of
degenerate states related by the symmetries of the Hamiltonian.}
\begin{equation}\label{A_ideal}
\boldsymbol{\phi}^{(A)}_k=(-1)^k
\left( \begin{array}{*{20}{c}}
  0&0&\lambda' \\ 
  0&0&0 \\ 
  0&0&0 \\
\end{array} \right)\equiv
(-1)^k
\left( \begin{array}{c}
  \boldsymbol{\eta} \\ 
  0 \\ 
  0 \\
\end{array} \right)
\end{equation}
with real $\lambda'$ and $(-1)^k$ indicating opposite sign on the two sublattices. While there are no ordered spin moments (dipoles) in this state, there is
a finite local spin density (see Fig.~{\ref{fig:laco1}) around Co atoms, which lead Halperin and Rice~\cite{halperin68} to introduce the name spin-density wave exciton condensate. 
Applying Hamiltonian symmetries one can generate all possible states belonging to the A-phase. In a general form, the A-type order parameter
has two rows of zeros and one row with real vector of the length $|\boldsymbol{\eta}|=\lambda'$.  We have repeated calculations with condensate in each of the three orbital channels ($\hat{x}$, $\hat{y}$, $\hat{z}$) as well as several general orientations of the spin-vector $\boldsymbol{\eta}$ to verify that these solutions are connected 
by the corresponding symmetry operations and lead to the same total energies. 
Given the difference of $\Gamma_{\alpha}$ matrices this is a numerically nontrivial test.

In the B-type state the condensate is distributed equally among all orbital channels. Similar to the A-type the order parameter is real and staggered
\begin{equation}\label{B_ideal_so3}
\boldsymbol{\phi}^{(B)}_k=(-1)^k
\left( \begin{array}{*{20}{c}}
  0&0&\lambda' \\ 
  0&0&\lambda' \\ 
  0&0&\lambda'\\
\end{array} \right)
\equiv
(-1)^k
\left( \begin{array}{c}
  \boldsymbol{\eta}_{\hat{x}} \\ 
  \boldsymbol{\eta}_{\hat{y}}  \\ 
  \boldsymbol{\eta}_{\hat{z}}  \\
\end{array} \right).
\end{equation}
The spin density in this phase, shown in Fig.~{\ref{fig:laco1}, is collinear and gives rise to no ordered dipole.
In the general form, the B-type order parameter consists real row vectors of equal length that are mutually parallel or anti-parallel,
$\pm\boldsymbol{\eta}_{\hat{x}}=\pm\boldsymbol{\eta}_{\hat{y}}=\pm\boldsymbol{\eta}_{\hat{z}}$. 

The C-type states have the most complicated structure. Similar to the B-type the condensate has equal weight in all orbital channels, 
but the corresponding spin vectors $\boldsymbol{\eta}_\alpha$ are mutually orthogonal. The dominant part of the $\phi$ is staggered and real, but there
is a small uniform and imaginary component, which has the same matrix structure as the staggered part
\begin{equation}\label{C_ideal}
\boldsymbol{\phi}^{(C)}_k=(-1)^k
\left( \begin{array}{*{20}{c}}
  \lambda'&0&0 \\ 
  0&\lambda'&0 \\ 
  0&0&\lambda' \\
\end{array} \right)
\begin{array}{*{20}{c}}
\end{array}
+
i\left( \begin{array}{*{20}{c}}
\lambda''&0&0 \\ 
  0&\lambda''&0 \\ 
  0&0&\lambda'' \\
\end{array} \right)
\begin{array}{*{20}{c}}
\end{array}.
\end{equation}
There is a complicated non-collinear distribution of spin density, shown in Fig.~{\ref{fig:laco1}, which does not give rise to an ordered dipole.
The general form of the $\boldsymbol{\phi}_k^{(C)}$ is $(-1)^k\lambda'+i\lambda''$ times an $O(3)$ matrix, where the $SO(3)$ part comes from the action of spin rotations  
on $\boldsymbol{\phi}_k^{(C)}$ of Eq.~\ref{C_ideal} and
$-\boldsymbol{\phi}_k^{(C)}$ is obtained by time reversal. The $O_h$ orbital symmetry does not add additional degenerate states in this case. 

For all studied values of the interaction parameters of $U$ and $J$ we found the $C$-phase to have the lowest energy among the stable
ordered solutions. In Table~\ref{table1} we compare the condensation energies (the difference between the total energy
of the ordered and normal state) for $U = 3.95$~eV and $J = 0.95$~eV.

% Already divided by 2! Formula unit corrected!
\begin{table}
\begin{tabular}{cccc}
     EC phase          & $| \boldsymbol{\phi} ' |$ & $| \boldsymbol{\phi} '' |$ & $E - E_0 [meV/f.u.]$\\
\hline
$A$-phase & 0.151 & 0.000 & -13.32  \\
$B$-phase & 0.191 & 0.000 & -20.39   \\
$C$-phase & 0.202 & 0.009 & -22.81   \\
\end{tabular}
\caption{The amplitudes of the staggered real $|\boldsymbol{\phi} ' |$ and uniform imaginary $|\boldsymbol{\phi} '' |$
components of the order parameter together with the condensation energies for the LDA+U solutions
($U$=3.90~eV, $J$=0.95~eV) in the cubic structure.}
\label{table1}
\end{table}

Stability of the excitonic condensate as a function of the interaction parameters is summarised in Fig.~\ref{phase_diag_UJ}.
We find that ordered solutions exist on the small $U$ side where the normal state is metallic as well as for large $U$ where the 
normal state has a well developed gap. At intermediate $U$ values we observed reduction of the order parameter $\boldsymbol{\phi}$ for larger $J$ values
and complete suppression of the ordered state for smaller $J$'s. This behavior is likely related to the crossover from weak coupling regime, where the transition 
involves mostly states close to the Fermi surface, to the strong-coupling regime, where the transition involves states in the entire Brillouin zone.
We find that the $A$-phase exists as a metastable solution only at intermediate $U$ and large $J$. Outside this region the $A$-like states converge to either
$B$ or $C$ solutions. The results in Fig.~\ref{phase_diag_UJ} allow for two general conclusions. Larger $J$ favors excitonic condensation,
most likely by bringing the excited IS states energetically closer the LS ground state. More stable state have larger order parameter $|\phi|$.

\subsection{Real LaCoO$_3$ structure without spin-orbit coupling}\label{section_real}
The actual structure of LaCoO$_3$ differs from the ideal cubic perovskite by a rhombohedral distortion. The crystal expands and the CoO$_6$ octahedra
rotate along one of the cubic body diagonals. In the following we study the impact of this structural modification on the excitonic phase. 
To this end we fix the interaction parameters at $U$= 3.90~eV and $J$=0.95~eV and construct several structures that continuously interpolate between 
the atomic positions of the cubic $\mathbf{r}_{cubic}$ and the real $\mathbf{r}_{real}$ structure
\begin{equation}
 \mathbf{r}= a  \mathbf{r}_{cubic} + (1 - a) \mathbf{r}_{real}.
\end{equation}
Upon lifting the cubic symmetry the anomalous part of the Co-$3d$ occupation matrix slightly deviates from (\ref{D_mm_ss}), nevertheless,
$\boldsymbol{\phi}$ can still be used as an order parameter.
%Nevertheless, the appearance of the non-zero spin-triplet elements still represents breaking of the spin rotational symmetry. The corrections to
%(\ref{D_mm_ss}) expressed in the coordinate system tied to CoO$_6$ octahedra are minor and thus we can still use (\ref{phi_expect_O}) as an 
%order parameter.
In Fig.~\ref{path_ideal_real} we show the evolution of the $B$ and $C$ phases with the distortion $a$. 
The distortion suppresses the excitonic LRO - no ordered solution
is found for the actual LaCoO$_3$ structure. For all $a$ values the $C$-phase retains the lowest energy among the stable states with LRO.

There are several more subtle observations. Eliminating some of the orbital symmetries splits the $B$-type states into two distinct sets
depending on the signs of $\boldsymbol{\eta}_\alpha$:
$B_{\text{ppp}}$ connected to $+++$ state and $B_{\text{mpp}}$ connected to $-++$ state.
The latter is found to be more stable.
%, which has lower energy of the two.
The $\boldsymbol{\phi}^{(B)}$ remains purely real for all studied structures and with no ordered spin dipoles.

There is no branching of the $C$ solutions since they all can be connected by spin rotations or
time reversal. The uniform imaginary part of $\boldsymbol{\phi}^{(C)}$ is suppressed faster than the staggered real 
part. The angle $\Theta$ between the spin-vectors $\boldsymbol{\eta}_\alpha$ (we consider only the dominant real part of $\boldsymbol{\phi}^{(C)}$)
varies from 90$^\text{o}$ to 120$^\text{o}$, where it saturates for $a\simeq0.6$. For the states with $\Theta$ between these two limits
we observe a small ferromagnetic polarization $m$. These observations are summarised in Fig.~\ref{path_ideal_real}.

\subsection{Spin-orbit coupling}
The SOC breaks the spin-rotational symmetry and introduces coupling within the $t_{2g}$ subshell as well as 
between the $e_g$ and $t_{2g}$ orbitals. The latter can be viewed in the present context as a source field generating
$\phi^{\text{(SOC)}}_{\alpha\beta}=i\lambda \delta_{\alpha\beta}$. This raises several questions. Does excitonic LRO survive
in the presence of SOC, i.e., is there a symmetry distinction between the normal and ordered phase? 
How does SOC interact with the excitonic condensate?

We have performed calculations ($U$=3.90~eV, $J$=0.95~eV) in the cubic structure starting from several initial conditions, corresponding 
to the normal state and $B$ and $C$-type LROs.
In all cases the calculations converged to two distinct, but symmetry related, selfconsistent states~\endnote{Note that with SOC
the symmetry operations must be applied to spin (rows) and orbital (columns) indices simultaneously.}
\begin{equation}
\label{eq:soc}
\begin{split}
\boldsymbol{\phi}_k=
(-1)^k
&\begin{pmatrix}
  -0.12&0.00&0.00 \\ 
  0.00&0.06&0.09 \\ 
  0.00&-0.09&-0.06 \\
\end{pmatrix}\\
+&i\begin{pmatrix}
 0.06&0.00&0.00 \\ 
  0.00&0.06&0.00 \\ 
  0.00&0.00&0.06 \\
\end{pmatrix}. 
\end{split}
\end{equation}
The uniform imaginary part is essentially the same as in the normal state, i.e., it is generated
by SOC and does not represent a spontaneous symmetry breaking.
Finite real part $\boldsymbol{\phi}'$, however,  breaks the time-reversal symmetry 
and thus distinguishes the ordered and the normal states. Naively, one might have expected that SOC selects one out
of the continuum of degenerate C-type states (\ref{C_ideal}) by locking the $\boldsymbol{\eta}_\alpha$ spin-vectors 
to specific crystallographic directions without changing their relative magnitudes and angles. 
The solution, shown in (\ref{eq:soc}), is more complicated and the ordered state exhibits no residual symmetry.  
The ordered spin moment per Co in (\ref{eq:soc}) is less than 0.001~$\mu_B$.

Comparing the condensation energies in the cases with and without  SOC (see Table~\ref{tab:tab2}) we find
that SOC promotes the condensation. Indeed, with SOC we found a marginally stable ordered solution also
in the real LaCoO$_3$ structure where it did not exist without SOC. 
\begin{table}
\begin{tabular}{lccc}
 Structure           & $| \boldsymbol{\phi} ' |$ & $| \boldsymbol{\phi} '' |$ & $E - E_0 [\text{meV/f.u.}]$\\
\hline
cubic, no SOC & 0.202 & 0.009 & -22.81   \\
cubic with SOC      & 0.197 & 0.095 & -25.36   \\
real with SOC      & 0.014 & 0.090 & -0.03   \\
\end{tabular}
\caption{\label{tab:tab2}
The amplitudes of the staggered real $|\boldsymbol{\phi}'|$ and uniform imaginary $|\boldsymbol{\phi}''|$ 
components of the order parameter together with the condensation energies for the LDA+U solutions
($U$=3.90~eV, $J$=0.95~eV). We compare the cubic structure (C-phase) with and without SOC and show the 
result for real LaCoO$_3$ structure with SOC (there is no ordered solution in real structure without SOC).
}
\end{table}

\section{Discussion}\label{sec:discussion}
To understand the behavior and stability of different excitonic phases is difficult and the all electron scheme, such as the Wien2k LDA+U used so far, 
is too complex for this purpose. Therefore we resort to a simplified model below. Before we do that let us summarize
the main trends observed in the numerical results: i) there is a preference for real staggered order, ii) condensate distributed among 
the orbital channels is more stable, iii) rhombohedral distortion suppresses the condensation,
iv) small imaginary uniform component of the order parameter appears in some phases,
v) the uniform component is suppressed fast than the staggered one by rhombohedral distortion, 
vi) the spin-orbit coupling favors formation of the condensate.  

To discuss these features in a comprehensible way we use an effective low energy model that lives
in a Hilbert space spanned by LS, IS and HS multiplets on each Co site. A further simplification
can be achieved in that we view the LS lattice as a vacuum state, IS state as an exciton~\cite{sotnikov16}
created by the bosonic operator $d^{\dagger}_{i\alpha\beta}$ with $i$, $\alpha=\hat{x},\hat{y},\hat{z}$ and $\beta=x,y,z$ being
the site, orbital channel and spin index, respectively. The HS state can be viewed as 
a bi-exciton consisting of two excitons with different orbital characters.~\endnote{Here we ignore
the fact that while $|\langle IS_{\hat{x}}|d_{\hat{x}}^{\dagger}|LS\rangle|=1$, adding
a second exciton, e.g., $|\langle HS_{\hat{z}}|d_{\hat{y}}^{\dagger}|IS_{\hat{x}}\rangle|\simeq\tfrac{\sqrt{3}}{2}$,
has a weight less than one. This is because only the hole $t_{2g}$, but not the electron $e_g$, components
of excitons in different orbital channels are mutually orthogonal.}
The resulting model is  a generalized version of the strong coupling model of Refs.~\onlinecite{balents00,kunes15}.
\begin{equation}
\label{eq:model}
\begin{split}
H&=\epsilon\sum_{i\alpha\beta}d^{\dagger}_{i\alpha\beta}d^{\phantom\dagger}_{i\alpha\beta}+\sum_{ij\alpha\beta}t_{ij}(\alpha)d^{\dagger}_{i\alpha\beta}
d^{\phantom\dagger}_{j\alpha\beta}\\
+&\frac{1}{2}\sum_{ij\alpha\beta}\bar{t}_{ij}(\alpha)(d^{\dagger}_{i\alpha\beta}d^{\dagger}_{j\alpha\beta}+h.c.)+H_3+H_4.
\end{split}
\end{equation}
Hard core constraint of maximum one boson in a given orbital channel is assumed.
The terms $H_3$ and $H_4$ contain three and four $d$ operators respectively and have a complicated form. The order
parameter $\boldsymbol{\phi}$ on a given site is simply the expectation value $\phi_{\alpha\beta}=\langle d_{\alpha\beta}\rangle$.

i) The preference for real staggered order is related to the positive signs of $t_{ij}(\alpha),\bar{t}_{ij}(\alpha)>0$ in (\ref{eq:model}) as can be seen by replacing 
the $d$ operators with their expectation values. The $t$-term favors staggered order irrespective of whether $\boldsymbol{\phi}$ is real or imaginary.
The $\bar{t}$-term is minimized by both staggered real and uniform imaginary $\boldsymbol{\phi}$. The staggered real order therefore mininizes both 
terms simultaneously. Positive $t_{ij}(\alpha)$ is a general feature of the perovskite structure arising from the relevant $t_{2g}\leftrightarrow t_{2g}$ and $e_{g}\leftrightarrow e_{g}$ 
nearest-neighbor hoppings having the same (negative) signs.~\cite{kunes14a} The $\bar{t}$-terms originate from inter-site cross-hopping
between  $t_{2g}\leftrightarrow e_{g}$ orbitals and the pair-hopping part of the on-site interaction.~\cite{balents00,kunes14a,kunes16}
Since the relevant cross-hoppings between nearest and next nearest neighbors in the cubic structure are forbidden by symmetry 
we are left with pair-hopping which yields $\bar{t}_{ij}(\alpha)>0$.~\cite{kunes14a}

ii) In the mean-field approximation the bilinear part of (\ref{eq:model}) has a high $O(9)$ symmetry (upon summation over the nearest neighbors).
 The matrix form of the order parameter $\boldsymbol{\phi}$ is thus determined by 
the $H_3$ and $H_4$ terms. The discussion of all symmetry allowed $H_3$ and $H_4$ terms is beyond the scope of this paper. We point out
that already in the case of single orbital flavor~\cite{kunes15} there are several contributions describing the inter-site repulsion and spin-exchange
between the bosons. Without detailed knowledge of the model parameters the following discussion is necessarily speculative. We propose the
nearest-neighbor $d$-$d$ repulsion to favor the distribution of condensate among the orbital channels. We argue that the $d$-$d$ repulsion 
between the $d$-bosons of the same orbital flavor is substantially stronger than inter-flavor repulsion. If so, a condensate distributed among
as many channels as possible generates less positive interaction energy than a condensate concentrated in a single channel.

The difference between inter- and intra-channel repulsion can be understood as follows. A single exciton can lower its energy by virtual
hopping to the empty neighbor sites. These processes are Pauli-blocked (hard-core constraint in Eq.~\ref{eq:model}) when neighbors are occupied by excitons of the same orbital 
flavor, resulting in $d$-$d$ repulsion. The virtual hopping is modified, but not blocked, by an exciton with different orbital flavor and thus the repulsion 
is weaker in this case.~\endnote{An inter-site attraction between different orbital flavors is possible in principle.}

iii) Suppression of the excitonic order by rhombohedral distortion may be a simple consequence of reduced hopping due to the Co-O-Co
bond bending. Since we do not have a detailed understanding of the detailed balance between the hopping channels that are reduced
and those that are opened due to the distortion, we leave this as a numerical observation.

The remaining points are discussed in the following section.

\subsection{Uniform imaginary order}
To understand the origin of the uniform component of the order parameter would require detailed knowledge of the
parameters in (\ref{eq:model}). Possibility of detailed quantitative mapping of the LDA+U results on model (\ref{eq:model})
is questionable and we do not have such ambition. We only demonstrate a possible mechanism
to obtain a finite uniform component and discuss its general features.
To this end we drop the orbital degeneracy of (\ref{eq:model}) and use a specific form of $H_4$, corresponding 
to spin-spin exchange~\cite{kunes15}
\begin{equation}
\begin{split}
H'=&\epsilon\sum_{i}\bd^{\dagger}_{i}\cdot\bd^{\phantom\dagger}_{i}+t\sum_{\langle ij\rangle}(\bd^{\dagger}_{i}\cdot\bd^{\phantom\dagger}_{j}+h.c.)\\
+&\bar{t}\sum_{\langle ij\rangle}(\bd^{\dagger}_{i}\cdot\bd^{\dagger}_{j}+h.c.)-J\sum_{\langle ij\rangle}(\bd^{\dagger}_{i}\wedge\bd^{\phantom\dagger}_i)\cdot
(\bd^{\dagger}_{j}\wedge\bd^{\phantom\dagger}_j).
\end{split}
\end{equation}
We will use the variational ground state with staggered real and uniform imaginary, mutually orthogonal, components
of the condensate
\begin{equation}
|\Psi\rangle=\prod_k (s|0\rangle_k+(-1)^k x|x\rangle_k+iy|y\rangle_k) 
\end{equation}
with the normalisation $s^2+x^2+y^2=1$. The variational energy per lattice site
is 
\begin{equation}
\label{eq:var}
\begin{split}
&\langle\Psi|H'|\Psi\rangle/N=\epsilon(x^2+y^2)\\
&-Z(\bar{t}+t)x^2(1-x^2)-Z(\bar{t}-t)y^2(1-y^2)\\
&+2Zx^2y^2(\bar{t}-J),
\end{split}
\end{equation}
where $Z$ is the number of nearest neighbors. 
Expression (\ref{eq:var}) is then minimized subject to the conditions
$x^2>0$, $y^2>0$ and $x^2+y^2\le1$.

Let us consider separately the effect of each inter-site term.
Both $t$ and $\bar{t}$ terms contribute to the condensation characterised by finite $|\phi|^2=(1-x^2-y^2)(x^2+y^2)$. 
The positive hopping $t>0$ contribution is minimized by staggered order irrespective of the complex 
phase of $\phi$, i.e. $y=0$ in (\ref{eq:var}). The positive pair creation $\bar{t}>0$ term has degenerate minima for staggered real 
or uniform imaginary $\phi$, i.e. solutions with the same $x^2+y^2$ are degenerate. The antiferromagnetic exchange term
$J>0$ favors states with equal $x^2=y^2$. The ratio $y/x$ is thus determined by the balance between the hopping $t$ and exchange $J$.

Tuning the system from pure LS state ($x^2+y^2=0$), e.g. by varying $\epsilon$, we first obtain the staggered condensate and only at
larger values of $x^2+y^2$ the uniform component possibly appears. This is because $t$ contributes to (\ref{eq:var}) in the second order 
while $J$ only in the forth order.

Let us discuss the relevance of the present toy model to the studied material. The hopping $t$ and pair creation $\bar{t}$ terms
in (\ref{eq:model}) are diagonal in orbital indices which can justify the reduction to a single orbital flavor. On the other hand
the exchange term is certainly not the only force driving the appearance of the uniform order parameter. Should it be so, uniform
$\boldsymbol{\phi}''$ would have appeared also in the A and B phases. The uniform $\boldsymbol{\phi}''$ appearing only in the C phase,
indicates that the forth order term in the real material has a more complicated structure combining several orbital flavors. 

Nevertheless, the present toy model shows how the imaginary uniform component $\boldsymbol{\phi}''$ of the order parameter can appear. 
Its origin in a 4th order term explains why it disappears faster than the real staggered component $\boldsymbol{\phi}'$, as shown in Fig.~\ref{path_ideal_real}.
Finally, spontaneous appearance of uniform $\boldsymbol{\phi}''$ explains the stabilising effect of SOC. The simultaneous presence of
uniform $\boldsymbol{\phi}''$  and  staggered $\boldsymbol{\phi}'$ generates a negative ($H_3+H_4$)
contribution to the energy of the ordered state, while the $\boldsymbol{\phi}''$ alone generates a positive contribution
proportional to $t$. Without SOC, both of these terms contribute to the condensation energy and partially cancel 
each other out. 
%Without SOC this is partially compensated by a positive contribution of $\boldsymbol{\phi}''$
%to the bilinear $t$-term. Both of these terms are missing in the normal state and thus contribute to the condensation energy. 
With SOC, roughly the same $\boldsymbol{\phi}''$ appears in the normal and the ordered phases
and thus the above positive $t$-proportional term does not contribute to the condensation energy.

\section{Conclusions}\label{sec:conclusions}
We have used the LDA+U approach to investigate excitonic magnetism in prototypical cubic perovskite with the $d^6$ configuration of 
transition metal ion as well as in the real structure of LaCoO$_3$. We found that stable solutions with exciton condensate exist for realistic interaction
parameters, i.e., solution with energy lower than that of normal state with full Hamiltonian symmetry. We have confirmed the 
spin-triplet $S=1$  and orbital-triplet $T_{1g}$ character of the condensed excitons. We have introduced and calculated the corresponding
order parameters and identified the basic trends in relations to the rhombohedral distortion and spin-orbit coupling.
All stable exciton condensate solutions exhibited broken time-reversal symmetry, but vanishing ordered moments, which can be 
viewed as a ordered arrangement of higher magnetic multipoles on the lattice.
Among the stable excitonic solutions there is still a preference for equal distribution of the condensate among the orbital and spin
channels. The rhombohedral distortion of the real LaCoO$_3$ structure was found to be detrimental to the exciton condensation, while the spin-orbit
coupling is favorable for the condensate formation. 
Given the absence of an experimentally observed phase transitions in LaCoO$_3$,
our results suggest that the material is close to the excitonic instability and suggest ways how the instability can be approached. 

\acknowledgements
We thank A. Sotnikov, A. Kauch and A. Hariki for discussions. This work  has received funding from the European Research
Council (ERC) under the European Union's Horizon 2020 research and innovation programme (grant agreement No 646807). 

\section{Appendix}
The relationship between the cubic and spherical harmonics:
%In compounds with a cubic crystal field, like LaCoO$_{3}$ system, the energy levels of the $d$ shell are split into two degenerated sets which correspond with the $t_{2g}$ %(three-fold degenerated) and the $e_{g}$ (two-fold degenerated) irreducible representation of the symmetry group $O_h$. One can choose the $\{d_{xy}, d_{xz}, d_{yz}, %d_{x^2 - y^2}, d_{3z^2 -r^2}\}$ orbitals to be the orthonormal set of eigenfunctions which represent the $t_{2g}$ and $e_{g}$ states. This so-called cubic harmonic basis %can be expressed in terms of $l=2$ spherical harmonics as follows:
\begin{equation}\label{cubic_basis}
\begin{split}
d_{xy}  & = \frac{i}{\sqrt{2}} \left(Y^2_{-2} - Y^2_{2} \right) \\
d_{xz} & = \frac{1}{\sqrt{2}} \left( Y^2_{-1} - Y^2_{1} \right) \\
d_{yz} & = \frac{i}{\sqrt{2}} \left( - Y^2_{-1} - Y^2_{1} \right)\\
d_{x^2 - y^2} & = \frac{1}{\sqrt{2}} \left( Y^2_{-2} + Y^2_{2} \right) \\
d_{3z^2 -r^2} & = Y^2_{0}\\
\end{split}
\end{equation}
The $ d_{z^2 - x^2}$ and $d_{y^2 - z^2}$ orbitals in the above cubic harmonic basis.
\begin{equation}
\begin{split}
 d_{z^2 - x^2} & = \frac{1}{2} \left(\sqrt{3} d_{3z^2 - r^2} - d_{x^2 - y^2}\right) \\
 d_{y^2 - z^2} & =- \frac{1}{2} \left(\sqrt{3} d_{3z^2 - r^2} + d_{x^2 - y^2}\right) \\
\end{split}
\end{equation}
%We show in Eq.\ref{gamma_cubic} and Eq.\ref{gamma_spherical} the $5 \times 5$ matrices which encode the orbital part of the $T_{1g}$ excitons in the cubic and %spherical harmonic basis respectively.
The $\Gamma_\alpha$ matrices (\ref{O_operators}) in the $\{d_{xy}$, $d_{xz}$, $d_{yz}$, $d_{x^2-y^2}$, $d_{3z^2-r^2}\}$ basis of cubic harmonics:
\begin{equation}\label{gamma_cubic}
\begin{split}
 & {\Gamma'_{\hat{x}}=  \frac{1}{2} \left( {\begin{array}{*{20}{c}}
  0&0&0&0&0 \\ 
  0&0&0&0&0 \\
  0&0&0&-1&-\sqrt{3} \\ 
  0&0&-1&0&0 \\
  0&0&-\sqrt{3}&0&0 \\
\end{array}} \right)} \\ 
&  {\Gamma'_{\hat{y}} = \frac{1}{2}\left( {\begin{array}{*{20}{c}}
  0&0&0&0&0 \\ 
  0&0&0&-1&\sqrt{3} \\ 
  0&0&0&0&0 \\
  0&-1&0&0&0\\
  0&\sqrt{3}&0&0&0 \\
\end{array}} \right)} \\ 
 & {\Gamma'_{\hat{z}}= \left( {\begin{array}{*{20}{c}}
  0&0&0&1&0 \\ 
  0&0&0&0&0 \\ 
  0&0&0&0&0 \\
  1&0&0&0&0 \\
  0&0&0&0&0 \\
\end{array}} \right)} \\ 
 & {\Gamma''_{\hat{x}}=  \frac{i}{2} \left( {\begin{array}{*{20}{c}}
  0&0&0&0&0 \\ 
  0&0&0&0&0 \\
  0&0&0&-1&-\sqrt{3} \\ 
  0&0&1&0&0 \\
  0&0&\sqrt{3}&0&0 \\
\end{array}} \right)} \\ 
&  {\Gamma''_{\hat{y}}= \frac{i}{2} \left( {\begin{array}{*{20}{c}}
  0&0&0&0&0 \\ 
  0&0&0&-1&\sqrt{3} \\ 
  0&0&0&0&0 \\
  0&1&0&0&0\\
  0&-\sqrt{3}&0&0&0 \\
\end{array}} \right)} \\
 & {\Gamma''_{\hat{z}} = i \left( {\begin{array}{*{20}{c}}
  0&0&0&1&0 \\ 
  0&0&0&0&0 \\ 
  0&0&0&0&0 \\
  -1&0&0&0&0 \\
  0&0&0&0&0 \\
\end{array}} \right)} 
\end{split}
\end{equation}
The $\Gamma_\alpha$ matrices (\ref{O_operators}) in the $\{Y_2$, $Y_1$, $Y_0$, $Y_{-1}$, $Y_{-2}\}$ basis of spherical
 harmonics:
\begin{equation}\label{gamma_spherical}
\begin{split}
  &{\Gamma'_{\hat{x}} = \frac{i}{4} \left( {\begin{array}{*{20}{c}}
  0&-1&0&-1&0 \\ 
  1&0&\sqrt{6}&0&1 \\ 
  0&-\sqrt{6}&0&-\sqrt{6}&0 \\
  1&0&\sqrt{6}&0&1 \\
  0&-1&0&-1&0 \\
\end{array}} \right)} \\
 & {\Gamma'_{\hat{y}} = \frac{1}{4} \left( {\begin{array}{*{20}{c}}
  0&1&0&-1&0 \\ 
  1&0&-\sqrt{6}&0&1 \\ 
  0&-\sqrt{6}&0&\sqrt{6}&0 \\
  -1&0&\sqrt{6}&0&-1 \\
  0&1&0&-1&0 \\
\end{array}} \right)} \\ 
&  {\Gamma'_{\hat{z}} = i \left( {\begin{array}{*{20}{c}}
  0&0&0&0&-1 \\ 
  0&0&0&0&0 \\ 
  0&0&0&0&0 \\
  0&0&0&0&0 \\
  1&0&0&0&0 \\
\end{array}} \right)}  \\ 
&  {\Gamma''_{\hat{x}} =  -\frac{1}{4} \left( {\begin{array}{*{20}{c}}
  0&1&0&1&0 \\ 
 1&0&\sqrt{6}&0&1 \\ 
  0&\sqrt{6}&0&\sqrt{6}&0 \\
  1&0&\sqrt{6}&0&1 \\
  0&1&0&1&0 \\
\end{array}} \right)} \\ 
 & {\Gamma''_{\hat{y}} = \frac{i}{4} \left( {\begin{array}{*{20}{c}}
  0&-1&0&1&0 \\ 
  1&0&-\sqrt{6}&0&1\\ 
  0&\sqrt{6}&0&-\sqrt{6}&0 \\
  -1&0&\sqrt{6}&0&-1 \\
  0&-1&0&1&0 \\
\end{array}} \right)} \\ 
 & {\Gamma''_{\hat{z}} = \left( {\begin{array}{*{20}{c}}
  1&0&0&0&0 \\ 
  0&0&0&0&0 \\ 
  0&0&0&0&0 \\
  0&0&0&0&0 \\
  0&0&0&0&-1 \\
\end{array}} \right)} 
\end{split}
\end{equation}

\end{document}